\documentclass{camera}

\usepackage[centertags]{amsmath}
\allowdisplaybreaks[4]
\usepackage{amsbsy}
\usepackage{amsfonts}
\usepackage{amssymb}
\usepackage[dvips]{graphicx}
\usepackage[dvips]{hyperref}

\begin{document}

%
\title{THEORY OF NEUTRINO OSCILLATIONS}

%
\author{Carlo Giunti}

%
\organization{INFN, Sezione di Torino, and Dipartimento di Fisica Teorica,
Universit\`a di Torino,
Via P. Giuria 1, I--10125 Torino, Italy
\null
\vspace{1cm}
\null
\text{\textsf{Talk presented at IFAE 2003, Lecce, 23-26 April 2003}}
}

\maketitle

%

The standard theory of neutrino oscillations
has been formulated in the middle 70's
(see Ref.~\cite{Bilenky:1978nj})
on the basis of some assumptions that we are going to review critically.
The main assumptions are four:

\renewcommand{\labelenumi}{\theenumi}
\renewcommand{\theenumi}{(A\arabic{enumi})}
\begin{enumerate}
\item
\label{A1}
Neutrinos are ultrarelativistic particles.
\item
\label{A2}
Neutrinos produced or detected in
CC weak interaction processes
are described by the
flavor states
\begin{equation}
|\nu_{\alpha}\rangle
=
\sum_{k}
U_{{\alpha}k}^{*}
\,
|\nu_{k}\rangle
\,,
\label{001}
\end{equation}
where
$U$ is the unitary mixing matrix,
$\alpha=e,\mu,\tau$,
and
$|\nu_{k}\rangle$ is the state of a neutrino with mass $m_k$.
\item
\label{A3}
The propagation time $T$
is equal to the
source-detector distance $L$.
\item
\label{A4}
The massive neutrino states
$|\nu_{k}\rangle$
in Eq.~(\ref{001})
have the same momentum,
$p_k = p \simeq E$
(``equal momentum assumption''),
and different energies,
$
E_k
=
\sqrt{ p^2 + m_k^2 }
\simeq
E
+
m_k^2/2E
$,
where $E$ is the neutrino energy neglecting mass effects
and the approximations are valid for ultrarelativistic neutrinos.
\end{enumerate}

The assumption \ref{A1} is correct, because neutrino masses
are smaller than about one eV
(see Refs.~\cite{Bilenky:2002aw,hep-ph/0310238})
and only neutrinos with energy larger than about 100 keV
can be detected
(see the discussion in Ref.~\cite{Giunti:2002xg}).
As we will see, the ultrarelativistic character of neutrinos
implies the correctness of the assumptions
\ref{A2} and \ref{A3}
and the irrelevance of the assumption
\ref{A4},
which is not realistic.

In Ref.~\cite{Giunti:1992cb} it has been shown that
the assumption \ref{A2} is not exact,
because the amplitude of production and detection of the massive neutrino $\nu_k$
is not simply given by $U_{{\alpha}k}^{*}$
(see also Ref.~\cite{Giunti:2002xg}).
In the ultrarelativistic approximation
the characteristics of the production and detection processes
that depend on the neutrino mass can be neglected,
leading to a correct approximate description of flavor neutrinos
through the states (\ref{001})
(see also Ref.~\cite{Bilenkii:2001yh}).

The assumption \ref{A3} follows from the ultrarelativistic approximation,
because ultrarelativistic particles propagate almost at the velocity of light.
However,
in the standard theory of neutrino oscillations
massive neutrinos are treated as plane waves,
which are limitless in space and time.
In order to justify the assumption \ref{A3}
it is necessary to treat massive neutrinos as wave packets
\cite{Giunti-Kim-Lee-Whendo-91},
which are localized on the production process
at the production time and
propagate between the production and detection processes at a velocity
close to the velocity of light.
Such wave packet treatment
\cite{Giunti-Kim-Lee-Whendo-91,Giunti-Kim-Coherence-98,Giunti:2002xg,Giunti:2003ax}
yields the standard formula for the oscillation length.
In addition,
the different group velocities of different massive neutrinos
imply the existence of a coherence length for the oscillations,
beyond which the wave packets of different massive neutrinos
do not jointly overlap with the detection process
\cite{Nussinov:1976uw,Kiers:1996zj}.

The wave packet treatment of neutrino oscillations
is also necessary for a correct description
of the momentum and energy uncertainties
necessary for the coherent production and detection of
different massive neutrinos
\cite{Kayser:1981ye,Beuthe:2001rc,Giunti:2003ax},
whose interference generates the oscillations.

Let us discuss now the assumption \ref{A4},
which has been shown to be unrealistic in
Refs.~\cite{Giunti:2001kj,Giunti:2003ax}
on the basis of simple relativistic arguments.
Indeed, the relativistic transformation of energy and momentum
implies that the equal momentum assumption
cannot hold concurrently in different inertial systems.
On the other hand,
the probability of flavor neutrino oscillations is independent from the
inertial system adopted for its measurement,
because the neutrino flavor is measured by observing charged leptons whose character
is Lorentz invariant
(\textit{e.g.} an electron is seen as an electron in any system of reference).
Therefore,
the probability of neutrino oscillations
is Lorentz invariant
\cite{Giunti:2000kw,physics/0305122}
and must be derived in a covariant way.
In fact,
the oscillation probability has been derived
without special assumptions about the energies and momenta of
the different massive neutrino components
both in the plane wave approach
\cite{Winter:1981kj,Giunti:2000kw,Bilenkii:2001yh}
and in the wave packet treatment
\cite{Giunti-Kim-Lee-Whendo-91,Giunti-Kim-Coherence-98,Giunti:2001kj,Giunti:2003ax}.

Let us briefly describe the covariant derivation of the neutrino oscillation probability
in the plane wave approach, in which
the massive neutrino states in Eq.~(\ref{001})
evolve in space and time as plane waves:
\begin{equation}
|\nu_{k}(x,t)\rangle
=
e^{- i E_k t + i p_k x}
\,
|\nu_{k}\rangle
\,.
\label{002}
\end{equation}
Substituting Eq.~(\ref{002}) in Eq.~(\ref{001})
and expressing the $|\nu_{k}\rangle$ on the right-hand side in terms of flavor states
($
|\nu_{k}\rangle
=
\sum_{\beta=e,\mu,\tau}
U_{{\beta}k}^*
\,
|\nu_{\beta}\rangle
$),
we obtain
\begin{equation}
|\nu_{\alpha}(x,t)\rangle
=
\sum_{\beta=e,\mu,\tau}
\left(
\sum_{k}
U_{{\alpha}k}
\,
e^{- i E_k t + i p_k x}
\,
U_{{\beta}k}^*
\right)
|\nu_{\beta}\rangle
\,,
\label{003}
\end{equation}
which shows that
at a distance $x$ and after a time $t$
from the production of a neutrino with flavor $\alpha$
the neutrino is a superposition of
different flavors.
The probability of flavor transitions in space and time is given by
\begin{equation}
P_{\nu_{\alpha}\to\nu_{\beta}}(x,t)
=
\left|
\langle\nu_{\beta}|\nu_{\alpha}(x,t)\rangle
\right|^2
=
\left|
\sum_{k}
U_{{\alpha}k}
\,
e^{- i E_k t + i p_k x}
\,
U_{{\beta}k}^*
\right|^2
\,,
\label{004}
\end{equation}
which is manifestly Lorentz invariant.

Considering ultrarelativistic neutrinos,
we apply now the assumption \ref{A3},
$t \simeq x = L$,
where $L$ is the distance traveled by the neutrino between production and detection.
The phase in Eq.~(\ref{004}) becomes
\begin{equation}
E_k t - p_k x
\simeq
\left( E_k - p_k \right) L
=
\frac{ E_k^2 - p_k^2 }{ E_k + p_k } \, L
=
\frac{ m_k^2 }{ E_k + p_k } \, L
\simeq
\frac{ m_k^2 }{ 2 E } \, L
\,.
\label{005}
\end{equation}
It is important to notice that Eq.~(\ref{005})
shows that the phases of massive neutrinos relevant for the oscillations
are independent from any assumption on the energies and momenta
of different massive neutrinos,
as long as the relativistic dispersion relation
$ E_k^2 = p_k^2 + m_k^2 $
is satisfied.
This is why the standard derivation of the neutrino oscillation probability
gives the correct result,
in spite of the unrealistic equal momentum assumption \ref{A4}.

Using the phase in Eq.~(\ref{005}),
the oscillation probability as a function of the distance $L$
has the standard expression
\begin{equation}
P_{\nu_{\alpha}\to\nu_{\beta}}(L)
=
\sum_{k}
|U_{{\alpha}k}|^2
|U_{{\beta}k}|^2
+
2
\mathrm{Re}
\sum_{k>j}
U_{{\alpha}k}
U_{{\beta}k}^*
U_{{\alpha}j}^{*}
U_{{\beta}j}
\exp\!\left(
-i\frac{\Delta{m}^2_{kj} L}{2E}
\right)
\,,
\label{006}
\end{equation}
which depends on the squared-mass differences
$ \Delta{m}^2_{kj} = m_k^2 - m_j^2 $.
Let us notice that the expression (\ref{006})
is still Lorentz invariant,
as shown in Ref.~\cite{physics/0305122},
because $L$ is not the instantaneous source-detector distance
but the distance traveled by the neutrino between
production and detection.

In conclusion,
the standard expression for the
probability of neutrino oscillations is
robust and can be derived starting from realistic assumptions
in the plane wave approach
\cite{Winter:1981kj,Giunti:2000kw,Bilenkii:2001yh},
in a quantum mechanical wave packet treatment
\cite{Giunti-Kim-Lee-Whendo-91,Giunti-Kim-Coherence-98,Giunti:2001kj,Giunti:2003ax}
and in the framework of Quantum Field Theory
(see Ref.~\cite{Beuthe:2001rc} and references therein).




%
\end{document}